\documentclass[sigconf]{acmart}
\pdfoutput=1
\usepackage{listings}
\usepackage{tabularx}
\usepackage{tablefootnote}
\usepackage{booktabs}
\usepackage{cleveref}
\usepackage{soul}

\lstdefinelanguage{yml}{
  keywords={License, AuthorName, AuthorEmail, WebSite, SourceCode, IssueTracker, RepoType, Repo, Binaries, Builds, versionName, versionCode, commit, subdir, sudo, gradle, AllowedAPKSigningKeys, AutoUpdateMode, UpdateCheckMode, CurrentVersion, CurrentVersionCode}, keywordstyle=\color{blue}\bfseries,
  moredelim=[is][commentstyle]{||}{££}, identifierstyle=\color{black},
  sensitive=false,
  comment=[l]{\#},
  commentstyle=\color{gray}\ttfamily,
  stringstyle=\color{orange}\ttfamily,
  morestring=[b]',
  morestring=[b]"
}

\lstdefinestyle{acmstyleYAML}{
    language=yml,
    basicstyle=\ttfamily\footnotesize,
    keywordstyle=\bfseries\color{blue},
    commentstyle=\color{gray},
    stringstyle=\color{orange},
    breaklines=true,
    breakatwhitespace=false,
    showstringspaces=false,
    frame=single,
    captionpos=b,
    numbers=left,
    numberstyle=\tiny\color{gray},
    numbersep=5pt,
    xleftmargin=2.5em,
    framexleftmargin=2em,
    xleftmargin=2em,
    framexleftmargin=1.5em,
    literate={/}{{/\allowbreak}}1 {?}{{?\allowbreak}}1 {&}{{\&\allowbreak}}1 {_}{{\_\allowbreak}}1 {=}{{=\allowbreak}}1 {/}{{\slash}}1
}

\usepackage{microtype}
\DisableLigatures[-, <]{encoding = *, family = tt* }

\copyrightyear{2026}
\acmYear{2026}
\setcopyright{cc}
\setcctype{by}
\acmConference[ACM REP '26]{ACM Conference on Reproducibility and Replicability}{July 20--22, 2026}{Delft, Netherlands}
\acmBooktitle{ACM Conference on Reproducibility and Replicability (ACM REP '26), July 20--22, 2026, Delft, Netherlands}
\acmDOI{10.1145/3820002.3828593}
\acmISBN{979-8-4007-2778-8/2026/07}

\usepackage{siunitx}

\usepackage[textsize=tiny]{todonotes}

\begin{document}

\title{Understanding Build Reproducibility in the F-Droid Ecosystem}

\newcommand{\telecomaffil}{\affiliation{\institution{LTCI, Télécom Paris,\\ Institut Polytechnique de Paris}
    \city{Palaiseau}
    \country{France}
  }}

\author{Denise Nanni}
\email{denise.nanni@telecom-paris.fr}
\orcid{0009-0009-9064-7070}
\telecomaffil

\author{Julien Malka}
\email{julien.malka@telecom-paris.fr}
\orcid{0009-0008-9845-6300}
\telecomaffil

\author{Stefano Zacchiroli}
\email{stefano.zacchiroli@telecom-paris.fr}
\orcid{0000-0002-4576-136X}
\telecomaffil

\author{Théo Zimmermann}
\email{theo.zimmermann@telecom-paris.fr}
\orcid{0000-0002-3580-8806}
\telecomaffil

\author{Gabriele D'Angelo}
\email{g.dangelo@unibo.it}
\orcid{0000-0002-3690-6651}
\affiliation{\institution{University of Bologna}
  \city{Bologna}
  \country{Italy}
}

\begin{abstract}
The security of open source applications benefits considerably from the possibility of rebuilding their source and verifying the output.
F-Droid, a prominent distribution for open source Android applications, systematically rebuilds them from source and tests their bitwise reproducibility \emph{at app publishing time}.
However, F-Droid offers no guarantee that app reproducibility will continue to hold \emph{in the future}.
As software ecosystems evolve, reproducibility may degrade, with potential negative consequences for software preservation and security.

We present the first empirical study of build reproducibility in the F-Droid app ecosystem.
Analyzing historical reproducibility logs, we find that the overall bitwise reproducibility rate has been steadily increasing \emph{over time} (as new versions of apps are published).
We then evaluate how reproducibility holds \emph{in time} for fixed app versions, by attempting to rebuild \num{18904} app versions that F-Droid had previously confirmed bitwise reproducible, published between September 2018 and February 2026, achieving an 83\% rebuild success rate, and identify missing dependencies as the dominant cause of failure, accounting for 76\% of non-rebuildable cases.
Among successfully rebuilt apps, 94\% are also bitwise reproducible---i.e., they still yield bitwise identical artifacts upon rebuild.

Together, these results show that while bitwise reproducibility largely holds for apps that can be rebuilt, rebuildability itself is highly sensitive to temporal decay.
\end{abstract}

\keywords{Reproducible Builds,
  reproducibility,
  rebuildability,
  Android,
  F-Droid
}

\maketitle

\section{Introduction}\label{sec:introduction}

In order to reach its end users, a software component must first be built from its source code and then distributed to the devices on which it will run. In Free and Open Source Software (FOSS) ecosystems, these steps are typically performed by third parties known as \emph{software distributions}.
This model greatly simplifies software installation, as distributions provide binaries that users can easily download, install, and execute without requiring deep technical knowledge.
However, it also introduces a trust relationship between users and software distributions. In a general case, users have no practical way to verify that a distributed binary executable was indeed produced from the corresponding source code.
This concern is part of the broader problem of \emph{software supply chain security}, which has received increasing attention in recent years due to the security risks associated with complex dependency chains or build infrastructures and a number of high-profile attacks~\cite{solarwinds, xz-backdoor}.

\emph{Reproducible builds} have long been considered by both practitioners and researchers as a practical solution to this problem~\cite{lamb_reproducible_2022}. A software component is said to be \emph{bitwise reproducible} if independent parties can rebuild it from the same source code and obtain a bitwise identical binary.
This property enables third parties to verify that distributed binaries correspond exactly to their published sources, thereby eliminating the need to blindly trust the entities performing the build and distribution steps. Beyond supply chain verification, rebuildability and bitwise reproducibility also play an important role in long-term software preservation, as they enable software artifacts to be reconstructed and validated years after their original publication.

Despite these advantages, the adoption of reproducible builds in large-scale software ecosystems tooling to improve software supply chain security still faces practical challenges. Achieving bitwise reproducibility often requires careful engineering, as build processes, compilers, and packaging tools may introduce sources of non-determinism. Over the past decade, the Reproducible Builds project~\cite{noauthor_reproducible_2023} has worked to identify and eliminate such issues across the FOSS ecosystems. While previous research has extensively studied the technical challenges involved in achieving high bitwise reproducibility rates~\cite{benedetti_empirical_2025, bajaj_unreproducible_2023, malka:nix, sharma2025causes}, reproducibility is typically evaluated only at the time a package is built. In practice, however, software ecosystems evolve continuously: build environments, compilers, dependencies, and packaging infrastructures change over time. As a result, software that was once reproducible may later become non-rebuildable or fail reproducibility checks.
This temporal degradation of reproducibility has important implications for both software preservation and supply chain security, as when a software artifact previously known to be reproducible fails a reproducibility check, it may indicate a potential compromise in the software supply chain. Despite its practical importance, the long-term stability of rebuildability and reproducibility has received little attention in the literature.

In this work, we investigate this problem in the context of Android applications distributed through the F-Droid ecosystem.
F-Droid\footnote{\url{https://f-droid.org/}.} is a FOSS distribution (app store) for Android that builds applications directly from their source code. Founded in 2010, the project introduced automated builds early in its development, generating historical build data exploitable for research purposes. Applications are included in the repository only if they can be successfully rebuilt from source by F-Droid’s infrastructure.
Besides performing reproducibility verification at publication time, F-Droid has several times performed distribution-wide reproducibility checks of historical artifacts, yielding valuable data points to study how reproducibility decays in time.

We distinguish reproducibility \textit{over time} from reproducibility \textit{in time}, addressing the first by examining the reproducibility status of apps across their version history, and the second by determining whether reproducibility of a version holds as time passes.
Leveraging F-Droid historical metadata, we perform the first reproducibility study that complements historical reproducibility data with a large scale experimental study, by rebuilding \num{18904} historical versions of applications that were successfully tested for reproducibility between September 2018 and February 2026.
Our goal is to understand how reproducibility evolves and to identify the factors that cause previously reproducible software to become non-rebuildable or non-reproducible.

We address the following research questions:
\begin{itemize}

\item \textit{RQ1}: How does the rate of bitwise reproducibility of apps published in the F-Droid store (as computed at the time of publication) evolve over time?
  In the following, we refer to this aspect as \emph{reproducibility over time}.

\item \textit{RQ2}: Do F-Droid apps that were bitwise reproducible at a given (past) point in time remain bitwise reproducible later?
  We refer to this aspect as \emph{reproducibility in time}.

\end{itemize}

\paragraph{Results}

Our study shows that while the overall bitwise reproducibility rate in the F-Droid ecosystem has increased over time, rebuildability of historical applications degrades as build environments evolve. A large majority of successfully rebuilt applications remain bitwise reproducible, but missing or unavailable dependencies represent the dominant cause of rebuild failures.

\paragraph{Paper structure}
\Cref{sec:relatedwork} presents the related work.
\Cref{sec:background} provides the necessary background on Android applications, reproducible builds and F-Droid ecosystem.
The methodology of our research is described in \Cref{sec:methodology}, while the results are presented in \Cref{sec:results-data} and \Cref{sec:results-rebuild}.
Finally, we discuss the implications of our findings in \Cref{sec:discussion}, and the threats to validity in \Cref{sec:threatstovalidity}, then concluding in \Cref{sec:conclusions}.

\section{Related work}\label{sec:relatedwork}

To the best of our knowledge, this paper is the first to look at the evolution of build reproducibility in the context of F-Droid.
However, there are previous studies looking at forms of reproducibility in the Android ecosystem, and more generally in the Java ecosystem.
This is also the first paper which both analyzes historical reproducibility results from a distribution infrastructure and performs large-scale rebuilding to evaluate how reproducibility holds in time.
In this section, we put this work in the context of reproducibility research in the Android ecosystem, in the Java ecosystem and in the broader FOSS ecosystem.

\subsection{Reproducibility in the Android ecosystem}

Pöll and Roland~\cite{poll2022automating} conducted a study on the reproducibility of build artifacts derived from the Android Open Source Project.
In their work, they claim that perfect bitwise reproducibility is too hard to achieve in the current state of the Android ecosystem, and they propose a notion of ``accountable builds'', where legitimate deviations are permitted.
Our work focuses on strict bitwise reproducibility, but we measure it on a set of Android applications that are maintained by a group who has devoted efforts to achieve it, whereas the study of Pöll and Roland was focused in particular on firmware images.

\subsection{Reproducibility in the Java ecosystem}

In the broader Java ecosystem, there have been some efforts related to reproducibility of Maven packages through the Reproducible Central project~\cite{reproduciblecentral}.
Keshani et al.~\cite{keshani2024aroma} have explored methods to automatically make it possible to reproduce Maven packages, even when the authors of the package did not explicitly make efforts to achieve it.
More work has been done since then in the same direction~\cite{hassanshahi2025unlocking}.

To further improve reproducibility of Maven packages, Sharma et al.~\cite{sharma2025causes} evaluated \emph{canonicalization}, a technique that transforms the build output by normalizing the non-deterministic parts, which then increases the reproducibility rate of packages that would otherwise not be bitwise reproducible.
This technique is conceptually similar to the notion of ``accountable builds'' by Pöll and Roland~\cite{poll2022automating}.

\subsection{Reproducibility in the broader FOSS ecosystem}

The first efforts around bitwise reproducibility were focused on the Debian distribution, and led to the creation of the Reproducible Builds project~\cite{reproduciblebuilds}, which many other distributions and projects have joined, including F-Droid.

Bajaj et al.~\cite{bajaj_unreproducible_2023} have used historical data from the Debian distribution to analyze the causes of unreproducibility in their packages, and how they were fixed over time. We share with this work the methodology of analyzing historical data to understand the evolution of reproducibility, but we complement it with a large-scale rebuilding of historical versions of apps to evaluate how reproducibility holds in time.

Several recent studies have performed large-scale builds to evaluate reproducibility. This was the case of several of the above-mentioned studies in the Java ecosystem, but also of Randrianaina et al.~\cite{randrianaina_options_2024}, where the authors study the impact of configuration options on reproducibility in the Linux kernel, and, most notably, of Benedetti et al.~\cite{benedetti_empirical_2025}, where the authors rebuild a large set of packages from various language ecosystems. These studies, however, do not focus on the temporal degradation of reproducibility.

The only study that studied bitwise reproducibility in time is by Malka et al.~\cite{malka:nix}, where the authors rebuilt Nix packages from the period 2017--2023, compared them with historical build outputs, and observed good reproducibility results which increase over time.
However, the authors did not have historical reproducibility data to compare with, and so they cannot know for certain if the increase in reproducibility they observe is only due to a positive historical evolution of reproducibility in the Nix package repository, or if it is also the consequence of a temporal degradation of reproducibility.
The authors also analyze rebuildability, but in their case, they obtain a rebuildability rate which is always above 99\%, whereas in our case, we observe that rebuildability degrades much more over time, and that the main difficulty to preserve bitwise reproducibility in time for F-Droid is to be able to rebuild the app in the first place.
This is likely related to the properties of Nix which allow to better preserve the build environment in time, as was empirically tested in a previous study by the same authors~\cite{malka_reproducibility_2024}.

\section{Background}\label{sec:background}

\subsection{Android applications}
Android\footnote{\url{https://www.android.com/}.} is a widespread Linux-based operating system for mobile devices, deployed by various manufacturers across a highly heterogeneous hardware landscape.
It natively supports C/C++ code, but the most used programming languages in Android development are Java and Kotlin, with the latter becoming the new standard since 2019.
Apps can also be developed using cross-platform frameworks, such as Flutter (Dart), React Native (JavaScript), and others~\cite{android:topLanguages}.
When using external frameworks, the application is typically wrapped in a Java or native code layer that enables its execution on the device.
Development is primarily conducted by using tools that allow compiling, building, testing, and debugging the code.
Compiling Java-based code requires the use of the \textit{Java Development Kit} (JDK), while the \textit{Android Software Development Kit} (SDK) provides the tools necessary to package the application.

\paragraph{Build systems and dependency management}
Build systems automate and ease the pipeline by managing dependencies, compilation phases, and packaging.
The standard build system for Android is Gradle, which was used by 95\% of Android apps in 2022, according to Liu et al.~\cite{android:buildSystems}.
These systems are highly configurable, which introduces high variability in build tasks.

A primary function of \textit{modern} build systems is dependency management.
Applications often consist of more than 50\% third-party code~\cite{zhan2021_tplandroidapps}.
Consequently, the dependency graph can grow exceptionally large due to transitive dependencies.
Libraries can be specified using pinned versions, version ranges, or simply the \textit{latest} release and the build system is responsible for resolving and fetching these dependencies based on the provided requirements.
Dependency resolution works on the complete dependency graph, including transitive dependencies, and can encounter conflicts.
Dynamic conflict resolution policies can lead to inconsistent version selection, introducing a significant source of build non-determinism~\cite{dependencyresolution}.

\paragraph{Android artifacts}
In Android, compilation of an application results in a binary artifact, the \textit{Android Package} (APK).
The APK is used to distribute and install the app into the devices, and contains necessary information for the app execution.
At its core, an APK is a compressed archive containing the compiled code, the resources and assets of the app, and some metadata.
Its binary representation depends on packaging details such as file ordering and compression.

The distribution phase requires APKs to be \textit{signed}, a mandatory Android prerequisite for installation.
The APK signature is obtained by encrypting with asymmetric cryptography the hash digest of the unsigned artifact.
The signature and the corresponding certificate are attached to the APK, and they are used to verify and compare APKs.

\subsection{F-Droid ecosystem}

F-Droid is a community-maintained distribution of FOSS apps on Android, consisting of two components: a package manager (the F-Droid client app) and an app repository (a hosted collection of APKs).
Unlike commercial platforms such as the \textit{Google Play Store}, F-Droid focuses on providing a transparent, privacy-respecting environment that leaves users in full control of their devices.

As a third-party app store, it acts as an intermediary;
the client app fetches metadata from the repository, and manages the installation and lifecycle of apps on the device.
The F-Droid project ensures trust by rebuilding apps from their source code---rather than accepting pre-compiled binaries---and explicitly flagging ``anti-features'' such as non-free or tracking software.

\paragraph{F-Droid data}
Each app of F-Droid's repository is described by a YAML file in the metadata directory of \texttt{fdroiddata}, \Cref{lst:metadata_example} shows the general structure.
Attributes range from general app information to build-specific instructions.
The \texttt{Builds} key contains app versions: every release must have a version name, a unique version code in the app scope and a commit reference, that can be either a hash or a tag.
The build process of each release can be customized to suit more complex set-ups, for instance by defining bash commands to install additional specific dependencies, such as the required JDK.

\begin{lstlisting}[style=acmstyleYAML, caption={App metadata template}, label={lst:metadata_example}, float]
WebSite: https://example.f-droid.app
SourceCode: https://gitlab.com/APPLICATION_UPSTREAM/ExampleCom

RepoType: git
Repo: https://gitlab.com/APPLICATION_UPSTREAM/ExampleCom
Binaries: https://gitlab.com/APPLICATION_UPSTREAM/ExampleCom/-/releases/v

Builds:
  - versionName: '1.0'
    versionCode: 123
    # Use the commit hash which the App should be built from
    commit: d94b5f7ec7c6d7602c78a5e9b8a5b8c94d093eda
    # Where build.gradle is:
    subdir: app
    sudo:
      - echo "deb https://deb.debian.org/debian trixie main" > /etc/apt/sources.list.d/trixie.list
      - apt-get update
      - apt-get install -y -t trixie openjdk-21-jdk-headless
      - update-alternatives --auto java
    gradle:
      - yes

# The keys to be used for signing the APK
AllowedAPKSigningKeys: aaaaaaaaaaaaaaaaaaaa

\end{lstlisting}

\paragraph{F-Droid server}
App lifecycles are managed by \texttt{fdroidserver}.
The \texttt{build} command loads app metadata, applies version-specific configurations and performs the build.
Prior to compilation, the \textit{scanning} step enforces F-Droid policy by inspecting the source code and failing if requirements are unmet.
The checks cover the presence of pre-built libraries, proprietary or non-free code patterns, and the sources used to fetch dependencies.
For Gradle-based builds, the system is configured to use the \texttt{gradlew-fdroid} wrapper, a custom utility that ensures the use of specific Gradle versions, and checks their integrity by verifying the checksum.
A successful build yields the artifacts, specifically the APK and the build log.

The build can be performed either locally or on a virtualized build server using the \texttt{--server} flag.
When this flag is applied, the host machine starts a Vagrant virtual machine (the \textit{build server}) from a basebox, uploads the necessary data (e.g., metadata, libraries, and source code), and initiates the build.
The host uses the \texttt{--on-server} flag to notify the guest environment that the process is executing within an isolated, single-use VM.
F-Droid provides provisioning scripts that generate a Debian basebox, which serves as a standard, disposable environment for each build.
Because major Debian releases ship with a specific default Java version, updating the F-Droid basebox to a new Debian major release may alter the default Java environment.

At the time of writing, the latest F-Droid server release is version 2.4.3, which utilizes Debian 13 Trixie with support for Java 21.

\paragraph{Packages listing}
To list all available packages, F-Droid provides JSON-structured index files containing general and version-specific metadata.
The indexes are split between the main repository, which populates the mobile client and website by default, and an archive.
This archive serves as a legacy repository and must be explicitly enabled in the mobile client to be displayed.

\subsection{Reproducible builds}
The Reproducible Builds project aims to increase the adoption of bitwise reproducibility in the FOSS community.
Many recognized projects such as \textit{Debian}, \textit{NixOS} and \textit{F-Droid} are actively involved, advocating for the widespread availability of reproducible software.

The term ``reproducibility'' in the context of software builds refers to the property of being able to recreate a process or an artifact multiple times with identical outputs.
This concept applies both to the configuration of build environments and to the artifacts they produce.
In general, a build environment is reproducible when it is defined so that it can be deterministically recreated, with exactly the same configurations, components and variables.
Build reproducibility encompasses two related but distinct properties: \textit{rebuildability} and \textit{bitwise reproducibility}.
An app is rebuildable if it is possible to create the artifact, and it is bitwise reproducible if artifacts generated in independent build executions are bit-by-bit identical.
These properties are important for software preservation and security:
the former ensures software is preserved in time, enabling later auditing, while the latter increases trust in the software supply chain by means of third-party verification of distributed artifacts.
Meaningful verification requires consistent source code and deterministic build instructions to guarantee identical binaries across varying environments.

\paragraph{Reproducibility in F-Droid}
F-Droid supports reproducible builds through two mechanisms, depending on app metadata.

Apps explicitly marked as reproducible are verified during the build process.
This requires either a URL to the upstream binary artifact or the inclusion of signing keys and signature files, enabling an automatic comparison between the locally built artifact and the upstream version.
If reproducibility fails at this stage, the app is not published.

An independent verification server tests all apps post-publication, including those already verified during the build process.
This service rebuilds applications from source, compares the resulting artifacts with those distributed in the repository, and publishes reproducibility logs.
Although officially described as a work in progress, the verification server has been operational since 2018 and maintains an extensive history of reproducibility logs.

Verification results are distributed as JSON files, both as a global list per package and as individual test records.
Each test log includes the verification date, outcome, and any associated error messages.
For failed tests, a separate Diffoscope\footnote{\url{https://diffoscope.org/}.} report provides a detailed analysis of the differences.
Resource-intensive packages may be explicitly marked as \textit{ignored} and are excluded from verification.

\section{Methodology}\label{sec:methodology}

The two research questions introduced in \Cref{sec:introduction} study build reproducibility at different times: \emph{over time}, from historical data, and \emph{in time}, by rebuilding fixed package versions and re-testing their bitwise reproducibility. We run the latter study on a subset of the catalog only---the packages last known to be bitwise reproducible.

We refine the two questions into the following sub-questions, which the remainder of the paper answers in turn:
\begin{itemize}
    \item \textit{RQ1.1}: How does bitwise reproducibility evolve over time?
    \item \textit{RQ1.2}: Do new versions of apps remain reproducible?
    \item \textit{RQ2.1}: Are historically-bitwise-reproducible apps rebuildable today?
    \item \textit{RQ2.2}: What are the causes of rebuild failures?
    \item \textit{RQ2.3}: Are rebuilt apps yielding identical binaries compared to the historical builds?
\end{itemize}

\subsection{Data collection}
To create our dataset, we collected data from the F-Droid app catalog and reproducibility logs.

We downloaded the index files of the repository and the archive on 17 February 2026.
We flattened and merged both indexes into a comprehensive dataset containing package data such as version codes and publication dates.
We then enriched this dataset with metadata from \texttt{fdroiddata}, incorporating build-specific attributes for each version, such as the binaries URL.

We identified a discrepancy where 61 packages existed only in the metadata and 245 only in the index (accounting for 0.35\% of the total).
The metadata contains the full historical list of packages ever published, while the index contains only the subset that can still be retrieved from the app page.
The missing index entries correspond to packages that were never published to the website because of build failures or reproducibility issues; they remain listed in the metadata, marked as \textit{disabled}, but are not shown in the client.
Some apps have been renamed over time, leaving an entry in the index without any correspondence in the metadata.

F-Droid provides a list of reproducibility data as a JSON file on the verification server page, but we chose to crawl\footnote{Using the reference URL \url{https://verification.f-droid.org/<package-name>_<version-code>.apk.json}.} every version log individually, since we found that the index was missing some entries.
We built our dataset of reproducibility logs and integrated it into the previous dataset.

We found \num{5309} packages with duplicate reproducibility logs.
We removed the duplicates, keeping the earliest attempts for the historical analysis, to better represent the status in the past, and the most recent attempts for the build reproducibility experiment, to avoid considering packages that regressed (more details can be found in \Cref{subsec:rq1-1}).
The final dataset consisted of \num{80139} package versions.

Further refinement of the experiment's dataset involved filtering out packages with missing reproducibility data (\num{44967}) or marked as non-reproducible (\num{16263}), as well as removing those marked as \textit{disabled} (\num{12})---as this status implies they were not buildable by F-Droid in the first place---and packages whose metadata lacked the \texttt{RepoType} field (42), which is required to determine the appropriate version control system to retrieve the source code.
The rebuild dataset comprised \num{18904} package versions.

\subsection{Historical data analysis}

We analyzed historical trends in bitwise reproducibility across package versions. First, we assessed data availability by quantifying missing logs and their distribution. For packages with available data, we examined version histories and release dates to compute the reproducibility rate, defined as the ratio of reproducible packages to the total in each time slice. For each interval, we determine a package's status from its latest release, i.e., the version a typical user would rely on at that time.

To evaluate the persistence of bitwise reproducibility, we define reproducibility capacity as the number of packages that were (bitwise) reproducible at least once, and observed reproducibility as those reproducible at a given time; their difference represents regressions. We further analyzed reproducibility transitions, excluding missing data and considering only observed status changes.

We classified packages into six archetypes grouped by stability: (i) stable---status unchanged (always reproducible or never reproducible); (ii) semi-stable---at most two transitions (regressed, regressed but fixed, improved); and (iii) unstable---more than two transitions (volatile, with repeated oscillations).

\begin{figure*}
  \centering
  \includegraphics[width=\linewidth]{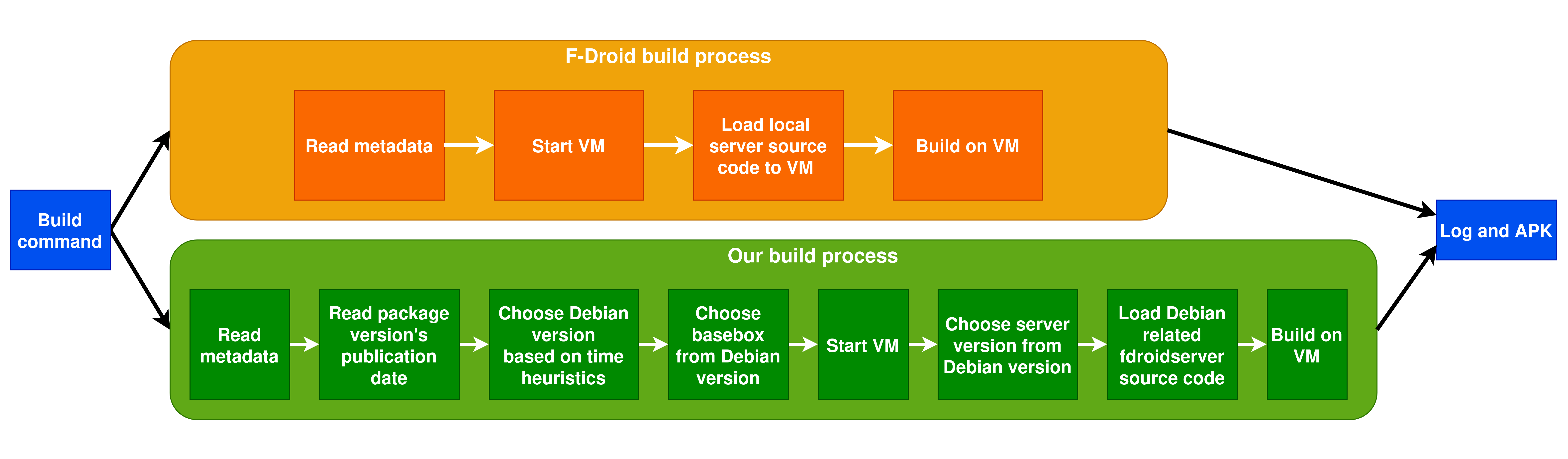}
  \caption{Comparison between F-Droid build process and our customization.}\label{fig:build-pipeline}
  \Description{}
\end{figure*}

\subsection{Rebuilding historical applications}
To answer \textit{RQ2.1}, we rebuilt apps from the F-Droid catalog to verify their rebuildability.
We relied on the VM-based configuration of F-Droid server, which ensures clean builds, and ran it on a cloud backend so that multiple builds could proceed in parallel.
Each app must be compiled in an environment as close as possible to the original one (including the Java version) to be meaningfully tested for reproducibility.

Achieving this required modifying both the F-Droid server source code and the virtual machine images, as the rest of this section describes: identifying the Debian basebox to use for each app, replicating the app legacy build environment, and aligning the build process with that environment.

\subsubsection{Debian version identification}
Each Debian release ships with a default JDK version, which is the one F-Droid uses for its builds. An incompatible JDK can cause build or reproducibility failures.
For this reason, it is important to select the correct basebox, which also determines the Java version. However, F-Droid keeps no explicit record of this: each app is simply built in the environment active when it was published.

F-Droid maintainers periodically update the Debian version used for the basebox.
According to the history of fdroidserver, in the considered time span, four different Debian releases were used: Stretch (9), Bullseye (11), Bookworm (12), and Trixie (13).

We therefore infer the basebox of each app from its publication date, using the time intervals shown in \Cref{tab:ostransitions}.
This ignores the gray areas around Debian transitions, but approximates the original environment well enough to rebuild a large number of apps without manually checking each one.

\begin{table}
    \centering
    \caption{Infrastructure operating system lifecycle with corresponding F-Droid server version tag.}
    \label{tab:ostransitions}
    \small
    \begin{tabularx}{\columnwidth}{llX}

        \toprule
        \textbf{OS version} & \textbf{Time period} & \textbf{Server tag} \\
        \midrule
        Debian 9 (Stretch)   & Prior to 03 Sep 2021 & 2.1.2 \\
        Debian 11 (Bullseye) & 03 Sep 2021 -- 13 Mar 2024 & 2.2.2 \\
        Debian 12 (Bookworm) & 14 Mar 2024 -- 17 Jan 2026 & 2.4.3 \\
        Debian 13 (Trixie)   & 18 Jan 2026 -- Present & 2.4.3\footnotemark \\
        \bottomrule
    \end{tabularx}
\end{table}

\footnotetext{For provisioning, we used revision bc3d22f4b2ba0dac595966e4d32acb23349ace37 since there was no explicitly tagged version.}

This heuristic occasionally fails to identify the correct version, in particular for packages published around the transition dates, which may have been built with different baseboxes due to processing delays.
Additionally, some apps' metadata specified custom commands that assumed certain configurations in the build server that were not compatible with the identified basebox.
In those edge cases, we manually checked the logs to identify the correct requirements and retried the build by explicitly setting the basebox to use. 

\subsubsection{Legacy environment replication}\label{subsub:envreplication}
Since we considered only packages with successful reproducibility logs, the oldest in our sample dates from 2018.
Recreating an environment this old is challenging: the source code may no longer be compatible with current tooling, and any substantial deviation from the original environment can break reproducibility.
We therefore provisioned one basebox per Debian release directly from the corresponding Debian image, using the server provisioning scripts mapped from versions in \Cref{tab:ostransitions}.
Adapting these scripts to our cloud backend required a few fixes: retaining packages that the minimal-image scripts purge but that our cloud VMs need, pulling packages from \url{archive.debian.org} for the archived Debian 9 and 11 releases (including working around expired certificates), and downloading the SDKs during provisioning on Debian 9, whose scripts assumed Vagrant shared folders, a feature unavailable in our cloud.

\subsubsection{Build process customization}
When invoked with \texttt{--server}, the build command delegates execution to a VM and, by default, uploads its own source code along with the required metadata and libraries.
This is problematic for legacy environments: the latest F-Droid server code does not even run on the Python 2 default of Debian 9.
To work around this, we cloned each server release separately and modified the build command to upload the version matching the identified Debian release (\Cref{tab:ostransitions}), so that the \texttt{fdroidserver} running in the VM matches its target environment.
\Cref{fig:build-pipeline} illustrates the differences between the F-Droid build process and our customization.

Two further adjustments let legacy servers accept current data.
Because the build command parses the entire, ever-growing metadata file before selecting a single version, newer fields or formats can make an old server reject an otherwise-valid historical build; we relaxed the parsers of legacy versions to tolerate these non-build-related changes.
We also handled the Gradle wrapper as each release expected, copying it from the matching server clone, since Debian 9 and 11 did not pre-provision it in the basebox.

\subsection{Build failures evaluation}
We determined build-failure causes by applying regular-expression heuristics to the log files.
We designed these expressions incrementally, inspecting samples of logs by hand and identifying common patterns for the most frequent causes of failure.
We then organized the causes into a taxonomy and used it to analyze the distribution of failures and the most common issues that arise when rebuilding historical applications.
We grouped the various errors into six categories.
Two categories refer to unavailability of software; in particular, \textit{missing dependencies} comprises all errors related to the unavailability of components, while \textit{missing source code} represents the unavailability of an app's source code.
The other categories focus on build errors: \textit{compilation errors} of any nature, \textit{component version mismatches}, which refer to failures caused by incompatible versions, and \textit{build environment errors}, which identify failures caused by limitations in the environment (e.g., process out of disk space or memory).
The \textit{other} category contains the remaining errors that do not fit into the other categories.

\subsection{Bitwise reproducibility verification}
We assessed bitwise reproducibility with the embedded verification where available, and tested the remaining packages with the \texttt{verify} command of the F-Droid server.

F-Droid server does not retain the APKs produced by failed builds, whether due to an actual build failure or a post-build issue (e.g., missing upstream binaries for reproducibility comparison).
We therefore modified the F-Droid server to retain the APK even when a build fails, to ensure that we could test all packages for bitwise reproducibility.

To further understand the causes of reproducibility failures, we generated Diffoscope reports and performed a qualitative analysis by inspecting the reports of 20 packages randomly sampled from the set of reproducibility failures.
We also analyzed a random sample of 20 Diffoscope outputs from the historical reproducibility logs to compare the causes of failures between the historical and rebuilt packages, and determine whether the same issues persist or if new ones arise upon rebuilds.

\section{Results: historical data analysis}\label{sec:results-data}
In this section, we present the results of the analysis of the historical data on reproducibility in F-Droid.

\subsection{RQ1.1: How does bitwise reproducibility evolve over time?}\label{subsec:rq1-1}
To understand the historical trend of reproducibility within the F-Droid repository, we first examined the annual distribution of reproducibility data.

\begin{figure}
  \centering
  \includegraphics[width=\linewidth]{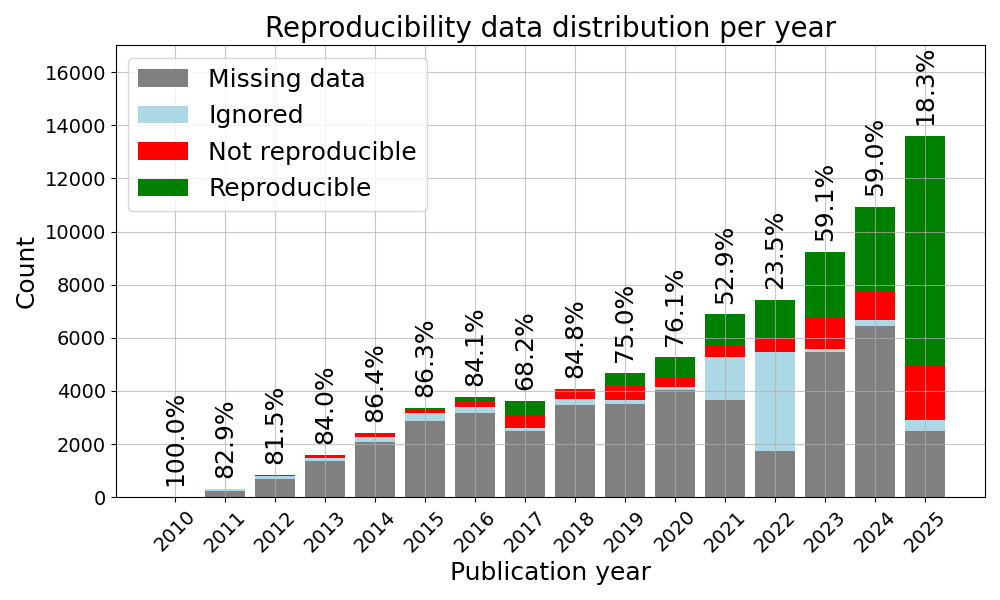}
  \caption{Yearly distribution of reproducibility data and the percentage of missing data.}\label{fig:missing-ignored-rep}
  \Description{Stacked bar chart that represents the total number of packages published per year. Grey area is the absolute number of packages with missing reproducibility data, light blue area are the packages with reproducibility test marked as ignored, red areas are the failed test and green areas are successful.}
\end{figure}

As illustrated in \Cref{fig:missing-ignored-rep}, while data gaps are prevalent in older packages, a sharp decline in missing data during 2021--2022 correlates with a simultaneous increase in ignored builds.
This suggests the verification server lacked the computational capacity and historical coverage of the primary build server in the past, while in 2025 it became able to cover the vast majority of the packages.

We then investigated the presence of duplicate reproducibility logs.
While recent duplicate attempts occurred close in time, \num{4878} packages published prior to 2021 registered a second attempt on a single day in 2021.
Because reproducibility tests systematically generated Diffoscope artifacts only from 2021 onward, we hypothesize that the entire repository was retested to retroactively provide these comparisons for all packages.

\begin{figure}
  \centering
  \includegraphics[width=\linewidth]{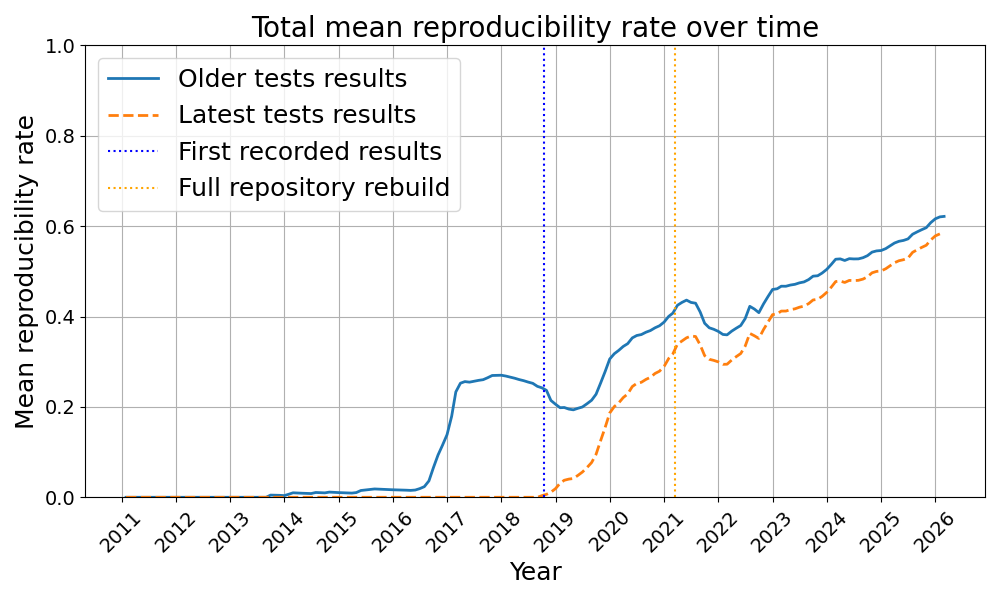}
  \caption{Monthly trend of reproducibility rate over time, comparing the results of the first and second attempts. For each month, reproducibility of an app is represented by its latest version's reproducibility result.}\label{fig:dup-rep-trend}
  \Description{}
\end{figure}

Notably, the two verification attempts yielded different results in many cases.
\Cref{fig:dup-rep-trend} compares the app reproducibility rates of the first and second attempts over time.
To compute these rates, we treated each month as a snapshot of the repository, evaluating the status of the latest release for each app;
if an app had no new release in a given month, its previous status was carried forward.

The upward slope observed between 2016 and 2018 can be attributed to two factors:
F-Droid's increasing effort to encourage reproducible builds, and temporal proximity.
Rebuilds executed closer to their original release are more likely to fetch identical dependencies and build environments, thereby achieving higher reproducibility rates.

Observing the results from the rebuild in 2021, it would seem that F-Droid unintentionally conducted a study on reproducibility in time.
According to the latest results, the rate of reproducibility dropped to zero for packages prior to the first quarter of 2018, suggesting that all applications three or more years old became unreproducible.
However, we suspect that this drop could be partly caused by changes in the verification criteria, particularly a tooling update, which may have introduced incompatibilities with older packages and caused them to incorrectly fail the reproducibility test.

\subsection{RQ1.2: Do new versions of apps remain reproducible?}
We analyzed the behavior of reproducibility across new versions of the same app to understand whether an app that is reproducible at a given time remains so across all its versions.

\begin{figure}
  \centering
  \includegraphics[width=\linewidth]{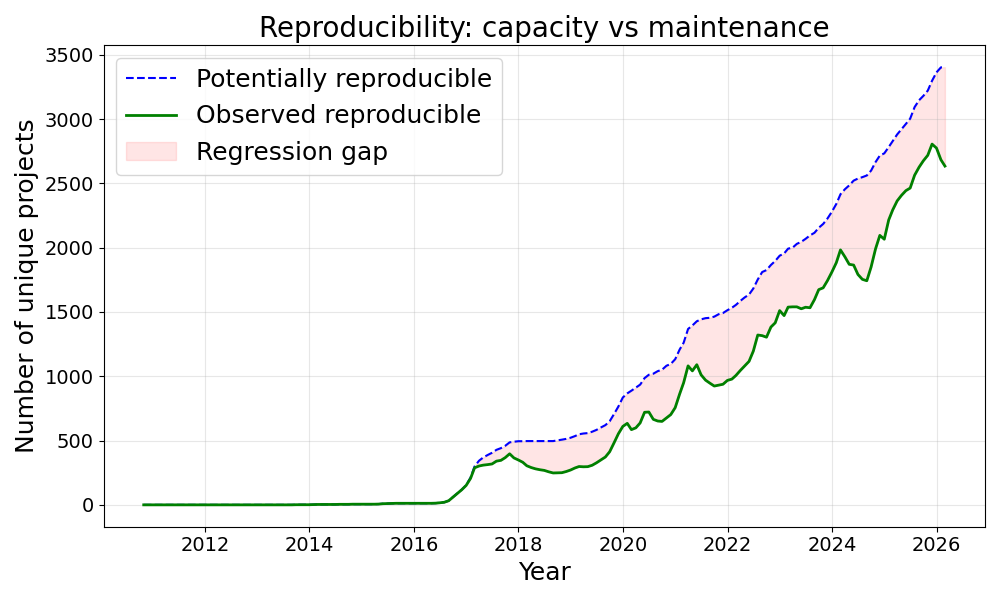}
  \caption{Monthly trend of reproducibility capacity versus observed status.}\label{fig:rep-cap-maint-trend}
  \Description{}
\end{figure}

\Cref{fig:rep-cap-maint-trend} shows the trend of reproducibility capacity compared to observed reproducibility over time, and highlights the regression gap.
We can observe a general positive trend in both metrics, with the gap tending to be quite stable overall.

\begin{figure}
  \centering
  \includegraphics[width=\linewidth]{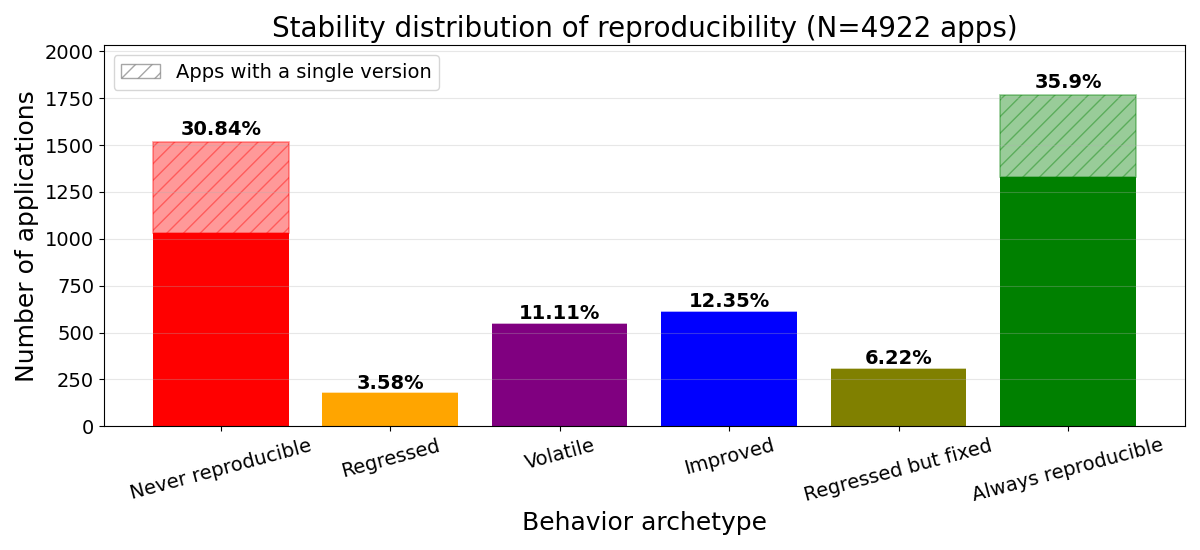}
  \caption{Archetype distribution of reproducibility stability.}\label{fig:archetype-distribution}
  \Description{}
\end{figure}

We also address the stability of reproducibility status by categorizing apps into archetypes, based on reproducibility transitions across new versions.
The distribution of apps in the different archetypes is shown in \Cref{fig:archetype-distribution}.
The majority of apps (66\%) are stable, i.e., they are either never or always reproducible, the latter being the most common archetype.
We observe a positive trend in the behavior of apps, with reproducibility fixes being more common than regressions.
Around 18.5\% of the apps have had reproducibility issues in their version history but are currently reproducible, while only 3.5\% regressed and are currently broken.
The remaining 11\% showed an unstable behavior, changing status multiple times in their lifetime.

\section{Results: build reproducibility in time}\label{sec:results-rebuild}
In this section, we answer the second subset of research questions by presenting the results of our package rebuilds and verification for bitwise reproducibility.

The sample we used to assess rebuildability consists of the packages that were historically bitwise reproducible, as the ultimate goal of rebuilding them is to assess whether they are still bitwise reproducible today.

\subsection{RQ2.1: Are historically-bitwise-reproducible apps rebuildable today?}
To assess current rebuildability, we attempted to rebuild each package version in our reproducible sample using the legacy-aware build pipeline described in \Cref{subsub:envreplication}.

Overall, we successfully rebuilt \num{15831} out of \num{18904} packages, corresponding to an 83.7\% rebuild success rate.
Since the sample is composed of package versions that were historically reproducible, these failures indicate that reproducibility in the past does not necessarily imply rebuildability in the present: even when the build recipe was once sufficient to reproduce the artifact, the build may later become impossible to complete due to ecosystem and infrastructure drift.

In the following, we break down the failures by their observed causes and quantify their prevalence.

\subsection{RQ2.2: What are the causes of build failures?}
These results have been achieved through multiple cycles of rebuilds, in which we identified and fixed external issues that were causing build failures.

The build process of apps with binaries in metadata includes a reproducibility verification step; in this case, F-Droid server treats a verification failure as a failed build.
This can occur either when the artifact is not bitwise reproducible or when the upstream binaries fail to be retrieved.
We consider those cases as successful for rebuildability purposes, because the build process is able to complete and the failure is related to bitwise reproducibility, which is the focus of the next research question.

We encountered build failures due to scanning errors that may have been caused by our pairing of \texttt{fdroidserver} source code with environments, because scanning policies have been updated over time, for instance narrowing the list of allowed source repositories for dependencies resolution.
The number of apps affected by this issue is limited---about 300 packages, accounting for around 1.5\% of the total---and we rebuilt them by relaxing the scanning policy.

Other build failures were caused by environment-related issues, such as a wrong Java version or Debian distribution.
There is a slight gray area around environment transitions that makes the time-based heuristic less accurate, and we tried to fix this issue by manually identifying the correct basebox for the affected packages.
We retried building 34 packages by explicitly setting the Debian version to use, and we were able to successfully rebuild 20 of them, most of which were published around the transition date from Debian 11 to Debian 12 and from Debian 12 to Debian 13.
The packages that still failed were mostly distributed across time with no clear correlation with the transition dates.

\begin{figure}
  \centering
  \includegraphics[width=\linewidth]{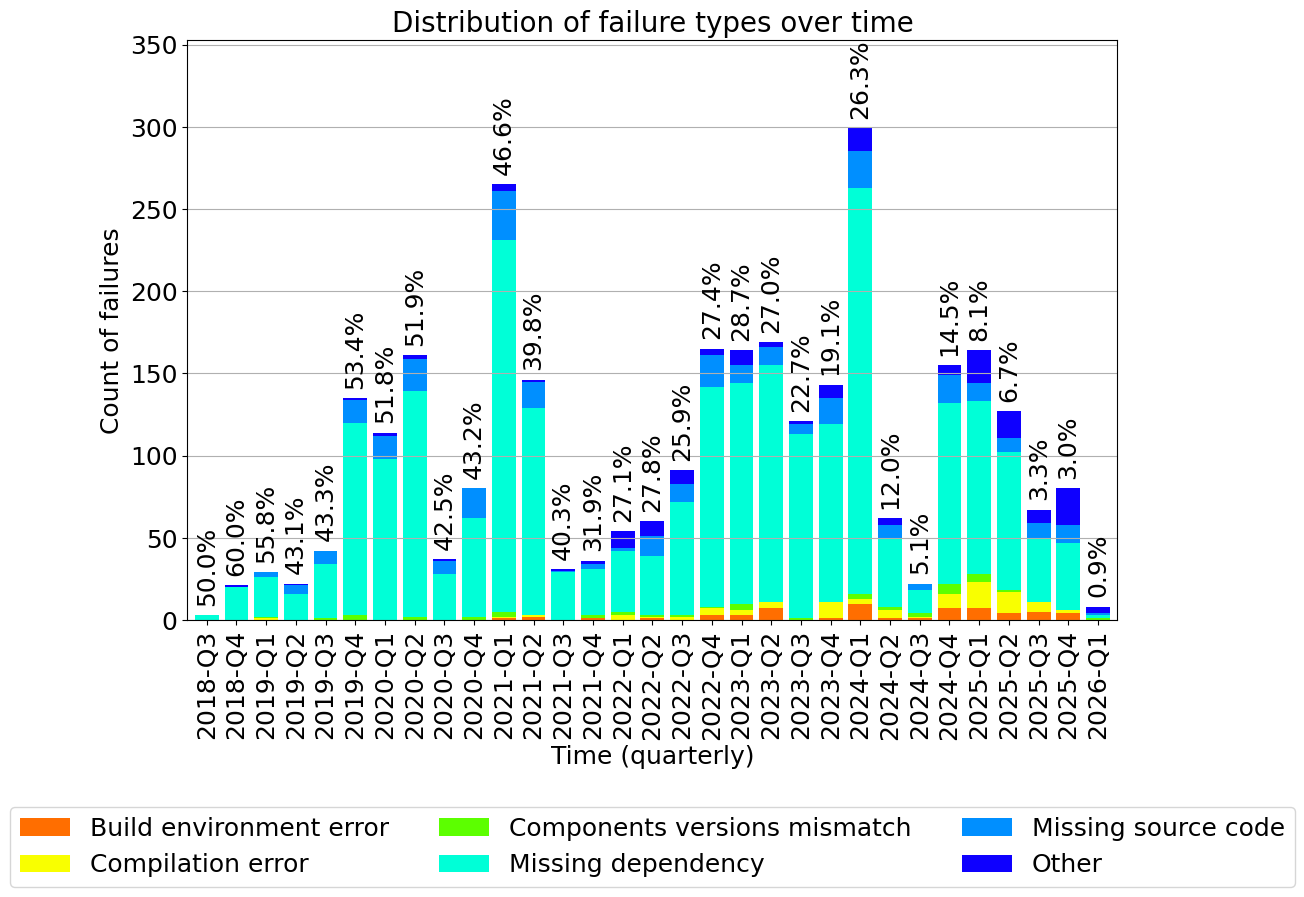}
  \caption{Quarterly distribution of build errors with percentage of rebuild failures in each interval.}\label{fig:errors-distribution}
  \Description{}
\end{figure}

\Cref{fig:errors-distribution} shows the distribution of build failures by cause, highlighting missing dependencies as the predominant cause, accounting for around 76\% of failed packages.
The second most common cause is the impossibility of retrieving the source code at all, accounting for 10\% of the failures.

Our methodology goes \emph{beyond} what a typical user would do, by using time-related heuristics to recover a build environment that is as close as possible to the original one, and by manually fixing edge cases.
This allows us to rebuild apps that would otherwise fail due to F-Droid's build tooling and policies in time.
However, fetching dependencies and source code is subject to third-party availability, which is outside our control.
Therefore, we consider these two dependency-related causes as true build failures and not as issues related to our methodology, because they reflect the reality of the difficulty of recovering the original build environment, a necessary condition to rebuild the app and test its reproducibility.

The percentages of failed rebuilds, shown in \Cref{fig:rep-ratio-results}, suggest that rebuildability degrades with the age of the package version.
Rebuild failures are generally more common for older releases; long-term rebuilds are more sensitive to the availability of external inputs than to the build process alone.

\subsection{RQ2.3: Are rebuilt apps yielding identical binaries compared to the historical builds?}
Among the rebuilt \num{15831} packages, \num{6348} were verified within the rebuild, while the remaining were separately tested with the \texttt{verify} command.
The embedded verification encountered \num{220} failures, accounting for 3.4\%.

Successfully built packages that failed because of missing upstream binaries (143 packages) were still candidates for reproducibility testing, since the unavailability of binaries does not mean they are not reproducible (and F-Droid has kept a copy of the binaries).
We tested them separately, obtaining a success rate of 90\%.

\begin{figure}
  \centering
  \includegraphics[width=\linewidth]{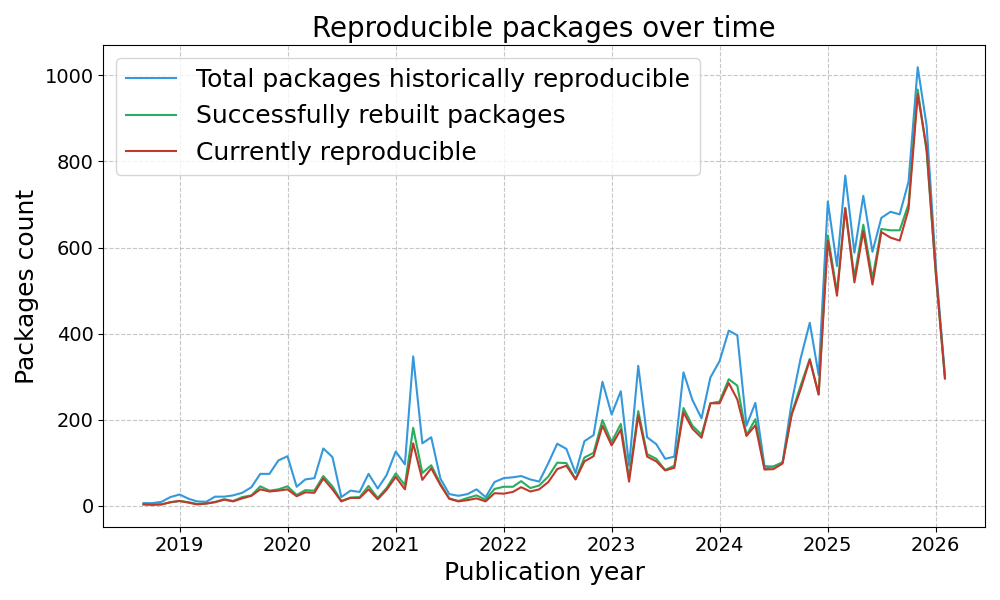}
  \caption{Number of packages with successful reproducibility test compared to historically reproducible and rebuilt packages.}\label{fig:rep-results}
  \Description{}
\end{figure}

\begin{figure}
  \centering
  \includegraphics[width=\linewidth]{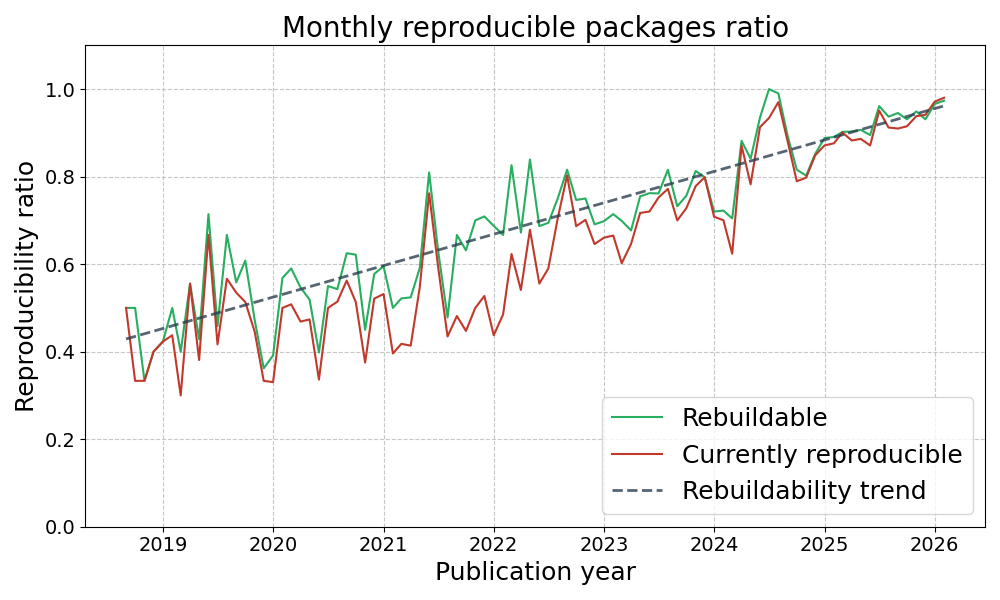}
  \caption{Comparison between the percentage of rebuilt packages and successfully verified packages within the reproducible sample. The dashed line illustrates the linear regression trend of the rebuildability curve (with $r = 0.9$).}\label{fig:rep-ratio-results}
  \Description{}
\end{figure}

Overall, we found 94\% of the successfully rebuilt package versions to still be bitwise reproducible.
\Cref{fig:rep-results} presents the distribution of historically reproducible, rebuilt, and currently reproducible packages, while \Cref{fig:rep-ratio-results} shows the rebuildability and reproducibility rates, computed against historically reproducible packages, with an overall positive trend as package publication dates are nearer to the rebuild date.
These charts confirm that temporal decay is a real concern for rebuildability;
however, the similarity of the reproducibility and rebuildability curves suggests that, once rebuildability is achieved, time has little effect on bitwise reproducibility.

\begin{figure*}
  \centering
  \includegraphics[width=\linewidth]{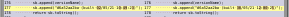}
  \caption{Historical Diffoscope output for \texttt{net.bible.android.activity:400} showing an embedded date in source code.}\label{fig:historical-diffoscope}
  \Description{}
\end{figure*}

The analysis of the Diffoscope outputs revealed that categories of current reproducibility failures slightly shifted from the historical ones.
In particular, embedded timestamps in the bytecode, a common cause of historical failures, did not appear in any of the current logs.
\Cref{fig:historical-diffoscope} provides an illustrative example of how such timestamp-related differences appear in a Diffoscope output.
Since our dataset comprises only packages that were historically bitwise reproducible, we can infer that this issue has been effectively addressed by F-Droid's verification process.

\begin{figure*}
  \centering
  \includegraphics[width=\linewidth]{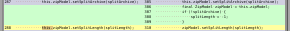}
  \caption{Generated Diffoscope report for \texttt{net.cozic.joplin:2097780} showing code differences.}\label{fig:joplin-diffoscope}
  \Description{}
\end{figure*}

Some of the remaining failures stem instead from dependency version mismatches, as in the case of \path{net.cozic.joplin:2097780}, and from embedded version strings, as in \path{com.github.premnirmal.tickerwidget:300900835}.
\Cref{fig:joplin-diffoscope} shows an excerpt of Joplin's Diffoscope report: the code addition is likely caused by a different version resolution for the library \texttt{zip4j}.
These failures are also present in the historical logs, although less frequently.

We also found cases in which the choice of JDK version directly affected bitwise reproducibility, as observed for \path{com.DartChecker:23} and \path{software.mdev.bookstracker:54}.
In these cases, the Diffoscope output reveals additional \textit{synthetic methods} in the bytecode, which are generated by the Java compiler and differ across compiler versions.

\section{Discussion}\label{sec:discussion}

Based on our experimental results, we can draw conclusions about the progress made by F-Droid in terms of bitwise reproducibility, and make recommendations on how to improve it further.

\subsection{Bitwise reproducibility progress in F-Droid}

The historical data analysis shows that the bitwise reproducibility rate in the F-Droid app ecosystem has steadily increased over time.
This may be due to internal efforts by F-Droid on their reproducibility infrastructure, as well as to the increasing awareness of this objective in the community.

The stability archetype analysis shows that app reproducibility, once achieved, is generally well-maintained: packages tend to gain reproducibility over time rather than regressing.
Given that we always consider the latest version of an app to evaluate the overall reproducibility rate of the F-Droid app ecosystem, this means that unmaintained apps contribute to a large extent to the remaining gap in reproducibility, preventing it from reaching a higher rate.

\subsection{Rebuildability as a foundation}

Our experimental findings show that bitwise reproducibility holds up in time \emph{provided we can successfully rebuild apps}, but this precondition is difficult to maintain.
Looking back five years, the number of ``reproducible'' apps that can still be tested is halved, primarily because rebuilding them has become harder.
Android builds depend on mutable external registries not designed for long-term archiving, as evidenced by frequent missing dependencies.
Moreover, many apps cannot be rebuilt because their source code is no longer available.

Unlike Nix, whose dependency model achieves rebuildability rates above 99\% even for packages several years old~\cite{malka:nix}, the F-Droid infrastructure inherits the weaknesses of the Android ecosystem.

\paragraph{Source code availability}

F-Droid recently began updating the source-code references for a subset of apps to \emph{point to archived versions} in Software Heritage~\cite{dicosmo:hal-01590958, f-droid-swh}.
While this practice has not yet been applied to build metadata, it signals growing awareness of the issue.
A concrete step toward preserving rebuildability is to \emph{archive build dependencies} alongside the source code and use these archives as fallbacks when the originals are no longer available from their hosts.

\textbf{Recommendation 1:} Developers should ensure that all source code required at build time---including that of the application being built and all its transitive dependencies---is (i) archived in a long-term repository such as Software Heritage and (ii) automatically retrieved from that archive if it becomes unavailable at the original hosting site.

\paragraph{The cost of under-specified environments}

Environmental drift is a major cause of non-rebuildability.
Packages built against specific component versions may fail once those components are updated, as with JDKs across Debian releases.
Replicating legacy environments and using time-based heuristics (like we did in our experiments) are pragmatic workarounds, but the root cause is that F-Droid does not systematically record the build environment at the time of publication.
Explicitly pinning the server version and the environment configuration in the build metadata or logs would improve long-term rebuildability and eliminate the gray areas we encountered during Debian transition periods.

Preserving all past build images would enable even higher rebuildability rates, since environmental drift can still cause failures even with a fixed Debian version.
If such an endeavor is infeasible for size or cost reasons, an alternative is to use VMs that can be rebuilt reproducibly, for instance by using a NixOS-based image, and by testing the bitwise reproducibility of the VM generation process.

\textbf{Recommendation 2:} document and ``pin'' the exact dependencies used and needed by all builds, for both apps and virtual machines.

Implementing recommendations (1) and (2) together will go a long way to address rebuildability issues, which we have identified as a major cause of non-bitwise reproducibility in the F-Droid app ecosystem.

\subsection{Security of the verification methods}

Reproducible builds in F-Droid already provide a positive security impact when verified during the build process.
In this mode, F-Droid checks the bitwise reproducibility of binaries produced and signed by upstream developers; agreement between these two independent entities on the resulting artifacts offers a strong guarantee to end users.

Post-publication verification, i.e., using F-Droid’s verification server to check the bitwise reproducibility of binaries produced by F-Droid itself, would provide strong additional guarantees only if the verification server were operated by an entity other than the one that builds the app.
This is an official goal of F-Droid’s work on the verification server~\cite{fdroid:verificationServer}, but independent entities have not yet deployed it.
Our experiments show that build reproduction is highly sensitive to environmental drift, which may undermine the effectiveness of third-party verification servers.
Providing a means to preserve the build environment in time would be a crucial step toward making it feasible for third parties to deploy their own verification servers and verify independent reproducibility, thereby providing stronger guarantees to end users.

\section{Threats to validity}\label{sec:threatstovalidity}

We adopt the terminology of Runeson et al.~\cite{runeson_guidelines_2009} in this section.

\paragraph{Construct validity} We reconstructed the original build environments with a best-effort policy, using time-based heuristics to infer which JDK version was used for each historical build. Better heuristics might yield higher rebuildability and reproducibility rates, but this does not undermine the insight that we infer from our results: that F-Droid's build environment specification is insufficiently precise to prevent temporal degradation of rebuildability and reproducibility.

Build failure causes were classified using regular expressions designed incrementally from manual log inspection. We did not formally validate their accuracy (e.g., through inter-rater agreement or a systematic audit of a random sample), so some failures may be misclassified.

\paragraph{External validity} Our study focuses on a single software distribution, F-Droid, and we do not claim that the specific reproducibility figures generalize beyond this ecosystem. In particular, tooling such as functional package managers has demonstrably better properties for preserving build environments in time. However, we have no reason to believe that our core insight---that build environment degradation is a major driver of reproducibility decay---would not hold in other ecosystems subject to similar environmental drift.

\paragraph{Reliability} We provide all the necessary code and data to reproduce our experiments. However, since a main takeaway of this paper is that build environment degradation impacts both rebuildability and, consequently, reproducibility, performing the same experiments later in the future would likely yield worse results.

\section{Conclusions}\label{sec:conclusions}

Achieving bitwise reproducibility at publication time is a crucial step for software supply chain security, but our study of the F-Droid ecosystem reveals that it is not sufficient for long-term preservation.
Although historical data show a positive trend in F-Droid's overall reproducibility rate, our effort to rebuild \num{18904} historically reproducible packages revealed significant temporal decay in rebuildability.
The fact that 94\% of successfully rebuilt packages retained their bitwise reproducibility shows that the primary bottleneck is not non-determinism in the build process itself, but rather environmental drift and the disappearance of source code and external dependencies.
Missing dependencies alone caused 76\% of the build failures in our study.
To mitigate this decay, software distribution infrastructures must evolve to archive all external dependencies alongside the source code and enforce rigorous pinning of historical build environments.

\section*{Data availability}
A complete reproducibility package for the work described in this paper is publicly archived and available from Zenodo~\cite{reproducibility-package}.

\begin{acks}
We thank the F-Droid project for openly maintaining the data and the infrastructure that made this research possible, and the maintainers who answered our questions.
\end{acks}

\clearpage



\begin{thebibliography}{24}


\ifx \showCODEN    \undefined \def \showCODEN     #1{\unskip}     \fi
\ifx \showISBNx    \undefined \def \showISBNx     #1{\unskip}     \fi
\ifx \showISBNxiii \undefined \def \showISBNxiii  #1{\unskip}     \fi
\ifx \showISSN     \undefined \def \showISSN      #1{\unskip}     \fi
\ifx \showLCCN     \undefined \def \showLCCN      #1{\unskip}     \fi
\ifx \shownote     \undefined \def \shownote      #1{#1}          \fi
\ifx \showarticletitle \undefined \def \showarticletitle #1{#1}   \fi
\ifx \showURL      \undefined \def \showURL       {\relax}        \fi
\providecommand\bibfield[2]{#2}
\providecommand\bibinfo[2]{#2}
\providecommand\natexlab[1]{#1}
\providecommand\showeprint[2][]{arXiv:#2}

\bibitem[noa(2023)]%
        {noauthor_reproducible_2023}
 \bibinfo{year}{2023}\natexlab{}.
\newblock \bibinfo{title}{Reproducible {Builds} — a set of software
  development practices that create an independently-verifiable path from
  source to binary code}.
\newblock
\urldef\tempurl%
\url{https://web.archive.org/web/20231113151826/https://reproducible-builds.org/}
\showURL{%
\tempurl}


\bibitem[Alkhadra et~al\mbox{.}(2021)]%
        {solarwinds}
\bibfield{author}{\bibinfo{person}{Rahaf Alkhadra}, \bibinfo{person}{Joud
  Abuzaid}, \bibinfo{person}{Mariam AlShammari}, {and}
  \bibinfo{person}{Nazeeruddin Mohammad}.} \bibinfo{year}{2021}\natexlab{}.
\newblock \showarticletitle{Solar Winds Hack: In-Depth Analysis and
  Countermeasures}. In \bibinfo{booktitle}{\emph{2021 12th International
  Conference on Computing Communication and Networking Technologies (ICCCNT)}}.
  \bibinfo{pages}{1--7}.
\newblock
\href{https://doi.org/10.1109/ICCCNT51525.2021.9579611}{doi:\nolinkurl{10.1109/ICCCNT51525.2021.9579611}}


\bibitem[Bajaj et~al\mbox{.}(2023)]%
        {bajaj_unreproducible_2023}
\bibfield{author}{\bibinfo{person}{Rahul Bajaj}, \bibinfo{person}{Eduardo
  Fernandes}, \bibinfo{person}{Bram Adams}, {and} \bibinfo{person}{Ahmed~E.
  Hassan}.} \bibinfo{year}{2023}\natexlab{}.
\newblock \showarticletitle{Unreproducible builds: time to fix, causes, and
  correlation with external ecosystem factors}.
\newblock \bibinfo{journal}{\emph{Empirical Software Engineering}}
  \bibinfo{volume}{29}, \bibinfo{number}{1} (\bibinfo{date}{Nov.}
  \bibinfo{year}{2023}), \bibinfo{pages}{11}.
\newblock
\showISSN{1573-7616}
\href{https://doi.org/10.1007/s10664-023-10399-4}{doi:\nolinkurl{10.1007/s10664-023-10399-4}}


\bibitem[Benedetti et~al\mbox{.}(2025)]%
        {benedetti_empirical_2025}
\bibfield{author}{\bibinfo{person}{Giacomo Benedetti}, \bibinfo{person}{Oreofe
  Solarin}, \bibinfo{person}{Courtney Miller}, \bibinfo{person}{Greg Tystahl},
  \bibinfo{person}{William Enck}, \bibinfo{person}{Christian Kästner},
  \bibinfo{person}{Alexandros Kapravelos}, \bibinfo{person}{Alessio Merlo},
  {and} \bibinfo{person}{Luca Verderame}.} \bibinfo{year}{2025}\natexlab{}.
\newblock \showarticletitle{An {Empirical} {Study} on {Reproducible}
  {Packaging} in {Open}-{Source} {Ecosystems}}.
\newblock


\bibitem[Di~Cosmo and Zacchiroli(2017)]%
        {dicosmo:hal-01590958}
\bibfield{author}{\bibinfo{person}{Roberto Di~Cosmo} {and}
  \bibinfo{person}{Stefano Zacchiroli}.} \bibinfo{year}{2017}\natexlab{}.
\newblock \showarticletitle{{Software Heritage: Why and How to Preserve
  Software Source Code}}. In \bibinfo{booktitle}{\emph{{iPRES 2017 - 14th
  International Conference on Digital Preservation}}}. \bibinfo{address}{Kyoto,
  Japan}, \bibinfo{pages}{1--10}.
\newblock
\urldef\tempurl%
\url{https://hal.science/hal-01590958}
\showURL{%
\tempurl}


\bibitem[F-Droid({[n.\,d.]})]%
        {fdroid:verificationServer}
F-Droid \bibinfo{year}{[n.\,d.]}\natexlab{}.
\newblock \bibinfo{title}{F-Droid - Verification Server}.
\newblock
\urldef\tempurl%
\url{https://f-droid.org/docs/Verification_Server/}
\showURL{%
Retrieved March 24, 2026 from \tempurl}


\bibitem[Hassanshahi et~al\mbox{.}(2025)]%
        {hassanshahi2025unlocking}
\bibfield{author}{\bibinfo{person}{Behnaz Hassanshahi},
  \bibinfo{person}{Trong~Nhan Mai}, \bibinfo{person}{Benjamin~Selwyn Smith},
  {and} \bibinfo{person}{Nicholas Allen}.} \bibinfo{year}{2025}\natexlab{}.
\newblock \showarticletitle{Unlocking Reproducibility: Automating re-Build
  Process for Open-Source Software}. In \bibinfo{booktitle}{\emph{2025 40th
  IEEE/ACM International Conference on Automated Software Engineering (ASE)}}.
  \bibinfo{pages}{3392--3402}.
\newblock
\href{https://doi.org/10.1109/ASE63991.2025.00280}{doi:\nolinkurl{10.1109/ASE63991.2025.00280}}


\bibitem[Keshani et~al\mbox{.}(2024)]%
        {keshani2024aroma}
\bibfield{author}{\bibinfo{person}{Mehdi Keshani},
  \bibinfo{person}{Tudor-Gabriel Velican}, \bibinfo{person}{Gideon Bot}, {and}
  \bibinfo{person}{Sebastian Proksch}.} \bibinfo{year}{2024}\natexlab{}.
\newblock \showarticletitle{AROMA: Automatic Reproduction of Maven Artifacts}.
\newblock \bibinfo{journal}{\emph{Proc. ACM Softw. Eng.}} \bibinfo{volume}{1},
  \bibinfo{number}{FSE}, Article \bibinfo{articleno}{38} (\bibinfo{date}{July}
  \bibinfo{year}{2024}), \bibinfo{numpages}{23}~pages.
\newblock
\href{https://doi.org/10.1145/3643764}{doi:\nolinkurl{10.1145/3643764}}


\bibitem[Lamb and Zacchiroli(2022)]%
        {lamb_reproducible_2022}
\bibfield{author}{\bibinfo{person}{Chris Lamb} {and} \bibinfo{person}{Stefano
  Zacchiroli}.} \bibinfo{year}{2022}\natexlab{}.
\newblock \showarticletitle{Reproducible {Builds}: {Increasing} the {Integrity}
  of {Software} {Supply} {Chains}}.
\newblock \bibinfo{journal}{\emph{IEEE Software}} \bibinfo{volume}{39},
  \bibinfo{number}{2} (\bibinfo{date}{March} \bibinfo{year}{2022}),
  \bibinfo{pages}{62--70}.
\newblock
\showISSN{1937-4194}
\href{https://doi.org/10.1109/MS.2021.3073045}{doi:\nolinkurl{10.1109/MS.2021.3073045}}


\bibitem[Liu et~al\mbox{.}(2024)]%
        {android:buildSystems}
\bibfield{author}{\bibinfo{person}{Pei Liu}, \bibinfo{person}{Li Li},
  \bibinfo{person}{Kui Liu}, \bibinfo{person}{Shane McIntosh}, {and}
  \bibinfo{person}{John Grundy}.} \bibinfo{year}{2024}\natexlab{}.
\newblock \showarticletitle{Understanding the quality and evolution of Android
  app build systems}.
\newblock \bibinfo{journal}{\emph{J. Softw. Evol. Process}}
  \bibinfo{volume}{36}, \bibinfo{number}{5} (\bibinfo{date}{April}
  \bibinfo{year}{2024}), \bibinfo{numpages}{20}~pages.
\newblock
\showISSN{2047-7473}
\href{https://doi.org/10.1002/smr.2602}{doi:\nolinkurl{10.1002/smr.2602}}


\bibitem[Malka et~al\mbox{.}(2024)]%
        {malka_reproducibility_2024}
\bibfield{author}{\bibinfo{person}{Julien Malka}, \bibinfo{person}{Stefano
  Zacchiroli}, {and} \bibinfo{person}{Théo Zimmermann}.}
  \bibinfo{year}{2024}\natexlab{}.
\newblock \showarticletitle{Reproducibility of {Build} {Environments} through
  {Space} and {Time}}. In \bibinfo{booktitle}{\emph{Proceedings of the 2024
  {ACM}/{IEEE} 44th {International} {Conference} on {Software} {Engineering}:
  {New} {Ideas} and {Emerging} {Results}}}
  \emph{(\bibinfo{series}{{ICSE}-{NIER}'24})}. \bibinfo{publisher}{Association
  for Computing Machinery}, \bibinfo{address}{New York, NY, USA},
  \bibinfo{pages}{97--101}.
\newblock
\showISBNx{9798400705007}
\href{https://doi.org/10.1145/3639476.3639767}{doi:\nolinkurl{10.1145/3639476.3639767}}


\bibitem[Malka et~al\mbox{.}(2025)]%
        {malka:nix}
\bibfield{author}{\bibinfo{person}{Julien Malka}, \bibinfo{person}{Stefano
  Zacchiroli}, {and} \bibinfo{person}{Th{\'e}o Zimmermann}.}
  \bibinfo{year}{2025}\natexlab{}.
\newblock \showarticletitle{{Does Functional Package Management Enable
  Reproducible Builds at Scale? Yes}}. In \bibinfo{booktitle}{\emph{{22nd
  International Conference on Mining Software Repositories}}}.
  \bibinfo{address}{Ottawa, Canada}.
\newblock
\urldef\tempurl%
\url{https://hal.science/hal-04913007}
\showURL{%
\tempurl}


\bibitem[Maven({[n.\,d.]})]%
        {reproduciblecentral}
Maven \bibinfo{year}{[n.\,d.]}\natexlab{}.
\newblock \bibinfo{title}{Reproducible Central}.
\newblock
\urldef\tempurl%
\url{https://github.com/jvm-repo-rebuild/reproducible-central}
\showURL{%
Retrieved March 13, 2026 from \tempurl}


\bibitem[Nanni et~al\mbox{.}(2026)]%
        {reproducibility-package}
\bibfield{author}{\bibinfo{person}{Denise Nanni}, \bibinfo{person}{Julien
  Malka}, \bibinfo{person}{Stefano Zacchiroli}, \bibinfo{person}{Th\'eo
  Zimmermann}, {and} \bibinfo{person}{Gabriele D'Angelo}.}
  \bibinfo{year}{2026}\natexlab{}.
\newblock \bibinfo{title}{Replication package for: Understanding Build
  Reproducibility in the F-Droid Ecosystem}.
\newblock
  \bibinfo{howpublished}{\url{https://doi.org/10.5281/zenodo.19217540}}.
\newblock
\newblock
\shownote{Last Accessed: March 22, 2026}.


\bibitem[Nesbitt(2026)]%
        {dependencyresolution}
\bibfield{author}{\bibinfo{person}{Andrew Nesbitt}.}
  \bibinfo{year}{2026}\natexlab{}.
\newblock \bibinfo{title}{Dependency Resolution Methods}.
\newblock
\urldef\tempurl%
\url{https://nesbitt.io/2026/02/06/dependency-resolution-methods.html}
\showURL{%
Retrieved March 20, 2026 from \tempurl}


\bibitem[Parthiban(2025)]%
        {android:topLanguages}
\bibfield{author}{\bibinfo{person}{Navin~Kumar Parthiban}.}
  \bibinfo{year}{2025}\natexlab{}.
\newblock \bibinfo{title}{Top 10 Best Programming Languages for Android App
  Development in 2025}.
\newblock
\urldef\tempurl%
\url{https://itechindia.co/blog/programming-languages-for-android-app-development/}
\showURL{%
Retrieved February 18, 2026 from \tempurl}


\bibitem[P\"{o}ll and Roland(2022)]%
        {poll2022automating}
\bibfield{author}{\bibinfo{person}{Manuel P\"{o}ll} {and}
  \bibinfo{person}{Michael Roland}.} \bibinfo{year}{2022}\natexlab{}.
\newblock \showarticletitle{Automating the Quantitative Analysis of
  Reproducibility for Build Artifacts derived from the Android Open Source
  Project}. In \bibinfo{booktitle}{\emph{Proceedings of the 15th ACM Conference
  on Security and Privacy in Wireless and Mobile Networks}} (San Antonio, TX,
  USA) \emph{(\bibinfo{series}{WiSec '22})}. \bibinfo{publisher}{Association
  for Computing Machinery}, \bibinfo{address}{New York, NY, USA},
  \bibinfo{pages}{6–19}.
\newblock
\showISBNx{9781450392167}
\href{https://doi.org/10.1145/3507657.3528537}{doi:\nolinkurl{10.1145/3507657.3528537}}


\bibitem[Randrianaina et~al\mbox{.}(2024)]%
        {randrianaina_options_2024}
\bibfield{author}{\bibinfo{person}{Georges~Aaron Randrianaina},
  \bibinfo{person}{Djamel~Eddine Khelladi}, \bibinfo{person}{Olivier Zendra},
  {and} \bibinfo{person}{Mathieu Acher}.} \bibinfo{year}{2024}\natexlab{}.
\newblock \showarticletitle{Options {Matter}: {Documenting} and {Fixing}
  {Non}-{Reproducible} {Builds} in {Highly}-{Configurable} {Systems}}. In
  \bibinfo{booktitle}{\emph{2024 {IEEE}/{ACM} 21st {International} {Conference}
  on {Mining} {Software} {Repositories} ({MSR})}}. \bibinfo{publisher}{IEEE},
  \bibinfo{pages}{654--664}.
\newblock
\urldef\tempurl%
\url{https://ieeexplore.ieee.org/abstract/document/10555868/}
\showURL{%
\tempurl}


\bibitem[Reproducible Builds Project({[n.\,d.]})]%
        {reproduciblebuilds}
Reproducible Builds Project \bibinfo{year}{[n.\,d.]}\natexlab{}.
\newblock \bibinfo{title}{Reproducible Builds Project}.
\newblock
\urldef\tempurl%
\url{https://reproducible-builds.org/}
\showURL{%
Retrieved March 13, 2026 from \tempurl}


\bibitem[Runeson and Höst(2009)]%
        {runeson_guidelines_2009}
\bibfield{author}{\bibinfo{person}{Per Runeson} {and} \bibinfo{person}{Martin
  Höst}.} \bibinfo{year}{2009}\natexlab{}.
\newblock \showarticletitle{Guidelines for conducting and reporting case study
  research in software engineering}.
\newblock \bibinfo{journal}{\emph{Empirical Software Engineering}}
  \bibinfo{volume}{14}, \bibinfo{number}{2} (\bibinfo{date}{April}
  \bibinfo{year}{2009}), \bibinfo{pages}{131--164}.
\newblock
\showISSN{1382-3256, 1573-7616}
\href{https://doi.org/10.1007/s10664-008-9102-8}{doi:\nolinkurl{10.1007/s10664-008-9102-8}}
\newblock
\shownote{Publisher: Springer Science and Business Media LLC}.


\bibitem[Sharma et~al\mbox{.}(2026)]%
        {sharma2025causes}
\bibfield{author}{\bibinfo{person}{Aman Sharma}, \bibinfo{person}{Benoit
  Baudry}, {and} \bibinfo{person}{Martin Monperrus}.}
  \bibinfo{year}{2026}\natexlab{}.
\newblock \showarticletitle{Causes and Canonicalization of Unreproducible
  Builds in Java}.
\newblock \bibinfo{journal}{\emph{IEEE Transactions on Software Engineering}}
  \bibinfo{volume}{52}, \bibinfo{number}{1} (\bibinfo{year}{2026}),
  \bibinfo{pages}{54--69}.
\newblock
\href{https://doi.org/10.1109/TSE.2025.3627891}{doi:\nolinkurl{10.1109/TSE.2025.3627891}}


\bibitem[Software Heritage Inclusion commit({[n.\,d.]})]%
        {f-droid-swh}
Software Heritage Inclusion commit \bibinfo{year}{[n.\,d.]}\natexlab{}.
\newblock \bibinfo{title}{Software Heritage Inclusion in F-Droid data}.
\newblock
\urldef\tempurl%
\url{https://gitlab.com/fdroid/fdroiddata/-/commit/b2082aee2232794be4aafcc85db314857e4f8cb0}
\showURL{%
Retrieved March 24, 2026 from \tempurl}


\bibitem[XZ backdoor({[n.\,d.]})]%
        {xz-backdoor}
XZ backdoor \bibinfo{year}{[n.\,d.]}\natexlab{}.
\newblock \bibinfo{title}{CVE-2024-3094}.
\newblock
\urldef\tempurl%
\url{https://nvd.nist.gov/vuln/detail/CVE-2024-3094?ref=thestack.technology}
\showURL{%
Retrieved March 24, 2026 from \tempurl}


\bibitem[Zhan et~al\mbox{.}(2022)]%
        {zhan2021_tplandroidapps}
\bibfield{author}{\bibinfo{person}{Xian Zhan}, \bibinfo{person}{Tianming Liu},
  \bibinfo{person}{Lingling Fan}, \bibinfo{person}{Li Li}, \bibinfo{person}{Sen
  Chen}, \bibinfo{person}{Xiapu Luo}, {and} \bibinfo{person}{Yang Liu}.}
  \bibinfo{year}{2022}\natexlab{}.
\newblock \showarticletitle{Research on Third-Party Libraries in Android Apps:
  A Taxonomy and Systematic Literature Review}.
\newblock \bibinfo{journal}{\emph{IEEE Transactions on Software Engineering}}
  \bibinfo{volume}{48}, \bibinfo{number}{10} (\bibinfo{year}{2022}),
  \bibinfo{pages}{4181--4213}.
\newblock
\href{https://doi.org/10.1109/TSE.2021.3114381}{doi:\nolinkurl{10.1109/TSE.2021.3114381}}


\end{thebibliography}
\end{document}